\documentstyle[12pt,aasms4]{article}

\def\cm2{~{\rm cm}^{-2}}

\def\nh0{N_{{\rm H}^0}}

\begin{document}

\title{Confirmation of high deuterium abundance in QSO absorbers}
\author{ M. Rugers and C. J. Hogan }
\affil{University of Washington}
\affil{ Astronomy, Box 351580, Seattle WA 98195-1580 }

\begin{abstract}

 We present   a new analysis of a Keck spectrum of
Q0014+813, with a new model of the $z=3.32$ absorbing cloud. We fit Lyman
series absorption of the dominant HI components, including
components identified with hydrogen and associated deuterium
at two discrete velocities, with thermal line broadening of each species.
The deuterium features are too narrow to be   interlopers,
and the good agreement in temperature and redshift with their hydrogen
counterparts confirms the identification as deuterium.
The abundance is measured to be $D/H$ $ = 1.9 \pm 0.5$ and $1.9 \pm 0.4
\times 10^{-4}$ in the two components, with an independent lower limit of
D/H $> 1.3 \times 10^{-4}$ for the sum, derived from the Lyman Limit
opacity. The impact on cosmological theory is briefly discussed.

\end{abstract}

\keywords{cosmology: observations ---   quasars: individual
 (0014+813)}

\section{Confirming Deuterium}

The hydrogen cloud at redshift $ z = 3.32 $ along the line of sight to the
quasar Q0014+813,   shown  by Chaffee et al (1985,86) to be uniquely
well suited for measuring  primordial abundances, was found recently to
show an absorption feature at the precise wavelength predicted for cosmic
deuterium (\cite{schr}, hereafter SCHR, and
\cite{car94}).  The implied D/H
abundance ratio from this feature, although it fits very well into the Big Bang
predictions for primordial abundances (\cite{cop95a},\cite{dar},\cite{fie96},
\cite{oli96}),
is  much higher than that usually derived from models of Galactic
chemical evolution
normalized by solar system $^3$He abundances (\cite{gal95}, \cite{wil94},
\cite{hat95},\cite{cop95b}), and also much higher than the best
estimate of $D/H$ in another QSO  absorber, on the line of sight to
Q1937-1009 (\cite{tyt94}). The simplest explanation of this discrepancy is that
the feature
in Q0014+0813 is not
caused by deuterium at all, but is simply a  low column density hydrogen
cloud that happens to lie at a somewhat different velocity  from
the bulk of the local hydrogen, at the   redshift expected for
deuterium (SCHR, and \cite{ste94}).  Although the redshift
coincidence is quite precise ($\pm 5$
km/sec), the chance  of
such a coincidence is still not  negligible, and was estimated to be at
least a few
percent, ignoring possible correlations.

We can eliminate the possibility of interlopers
 even in a single system if the
candidate D lines are very narrow. The linewidth, characterized by the
 Doppler parameter in a thermal profile fit,
$b \equiv \sqrt{{2kT}\over{m}}$
 is $  13 T_{4}^{1/2}{\rm km\ s^{-1}}$ for hydrogen,
$9 T_{4}^{1/2}{\rm km\ s^{-1}}$ for deuterium.
The   potential  interloper population is observed to have   lines below
$b=20 {\rm km\ s^{-1}}$ only very rarely (\cite{hu}), so a   narrower line is
much more naturally identified with deuterium than with a chance interloper.
Also
in very narrow lines, thermal broadening is likely to dominate the linewidth,
so the ratio of the fitted Doppler parameter for D candidates to their H
counterparts
is an important clue.
Here we use evidence from line shapes to confirm the identification of
the deuterium features and sharpen the estimate of the primordial deuterium
abundance.

\section{Observations and Reduction}

We use the same  data described in SCHR: six exposures
of 40 minutes each were obtained of Q0014+813, at the
Keck telescope on November 11, 1993, using the HIRES echelle spectrograph
with a $1.2" \times 3.8"$ slit. A resolution R=36,000 was obtained over the
covered wavelength range of 3500 to 6080 \AA. Unlike SCHR, the raw data
were reduced using IRAF. The CCD frames were overscan corrected, bias
subtracted and flat fielded using CCDPROC. After verifying the registration
of the frames, the frames were then stacked using IMSUM,
with rejection of the highest pixel value among the frames, in order to
reject cosmic rays. Fewer cosmic rays were left using this method than by
creating a median with IMCOM. The stacked images were then used for extraction
of
the individual orders of the spectrum using DOECSLIT, where the standard star
frame was used to trace the orders on the CCD and ThAr calibration lamp data
were used to obtain the wavelength calibration. No attempts were made to
flux calibrate the spectrum. The resulting extracted, dispersion corrected
echelle spectrum was then run through RVCORRECT to obtain the correction for
radial velocity of the observer with respect to the QSO, so redshifts quoted
here are in vacuo, relative to the local standard of rest. The data were not
smoothed, as they were in SCHR; the new results presented here
are due primarily to using the full instrumental resolution of the raw data.

\section{Model Fits for HI and  DI}

After examining the Lyman series of this absorber complex, we
decided to use Lyman-$\alpha$, $\beta$, $\gamma$, $\delta$, $\zeta$,
and $\kappa$ through 17 for establishing fits. Lyman-$\epsilon$ was thrown
out due to the presence of significant absorption from Ly$\alpha$-forest
interlopers. A value for $\log N_{HI}$ of 17.3 (estimated from the Lyman limit
optical depth) was used as an initial guideline for the column
density of the bluest part of the complex, where most of the $HI$ is located,
as is shown by the higher order lines. The code used to perform the model
fits to the spectrum is
VPFIT, which models the formation of an absorption spectrum as a series of
thermally broadened components at discrete velocities
(\cite{car87}, Webb 1987). For each component
VPFIT determines the redshift ($z$), Doppler parameter ($b$) and column density
($N$) by fitting the data with Voigt
profiles convolved with a Gaussian instrument profile, and makes formal error
estimates from the covariance matrix parameters, for each of the
calculated $z$, $b$ and $N$, based on the reduced
$\chi^{2}$. (Note that although these are ``true'' 1$\sigma$ errors,
including the total fitting uncertainty in each parameter, the probablility
distribution is highly non-Gaussian, so they do not translate directly into
``confidence intervals''. The results quoted in SCHR
were used as a starting point for the fits performed by VPFIT.

The Doppler parameter for the $D$-component fitted to
the smoothed data by SCHR, 14 km/s (assuming thermal broadening only),
is too large for a good fit to the unsmoothed data.
A narrower fit is required both by
the steep profile edge of the D feature at the blue side of this absorber
complex, and by a sharp spike in the center, which may at first sight appear
to be noise. However, since in the top part of this spike there are 3
channels, each of which has between
6 and 7 $\sigma$ counts above the  bottom of the absorption dip, this is
statistically very unlikely to be a noise feature. The feature also appears
if the same data are reduced using the independent software used by SCHR,
indicating it is not due to the different procedure performed here (A.\
Songaila, private communication). A much better fit
was obtained using two $HI$ absorbers instead of one,
each of which has a $\log N_{HI}$ of around 16.8, both with corresponding $DI$
absorption lines.
This two-component  fit is also preferred by the HI lines
on their own, without the split D$\alpha $ feature; by adding the three
new parameters
corresponding to the one new HI component,
the reduced $\chi^2$ per degree of freedom for the best fit
is equal to 0.98, instead of the
6.1 obtained for the best fit with only one main HI hydrogen absorber.
This is mainly due to the improved fit in the higher Lyman series lines.
[more here]

The   best overall fit was obtained by the
components as presented in Table 1, where the components of interest
for the deuterium abundance are numbers 11 to 14.
 Figure 1 shows plots
of the data and the fit to Ly-$\alpha$ through
$\delta$, plus $\zeta$, all plotted on the same velocity scale.
Most interloper features in the higher orders were omitted from
the fit.  Some areas of poorly fitted excess flux in $\delta$ and $\zeta$ must
be due to
artifacts or noise, as corresponding flux does not appear in higher orders.
The  highest
order line fits are shown in Figure 2.
Although many parameters of the minor components are not well determined by the
fit,
the HI columns of the dominant components (12 and 14)
are well constrained by the line fits in the
unsaturated high Lyman
series, and the DI columns are well-determined by the unsaturated Ly-$\alpha$
lines, so the absolute deuterium abundances  can be estimated fairly
accurately.

The $D/H$ ratios for both of the main
absorbers are in good agreement with each other, as well as with the value
obtained by SCHR. For the $z$ = 3.320482 absorber we find $(D/H) = 10 ^{-3.73
\pm 0.12}$, and for the $z$ = 3.320790 absorber we find $(D/H) = 10 ^{-3.72
\pm 0.09}$, where we have added the errors in $\log N_{HI}$ and $\log N_{DI}$
in quadrature.  We therefore expect the $D/H$ ratio for these absorbers to lie
within the ranges of $ 1.9_{-0.5}^{+0.6} \times 10^{-4} $ and
$ 1.9_{-0.4}^{+0.5}
\times 10^{-4}$, respectively.
 The total D column is $ 2.6 \pm 0.4 \times
10^{13}$, which also gives $ (D/H) = 1.9\pm{0.4} \times 10^{-4}$ when compared
to total hydrogen column in the two main absorbers. These are
consistent with the estimates made by SCHR and Carswell et al.\ (1994), using
lower resolution.

The total $N_{HI}$ in the two main absorbers from this fit is
$ 1.37 \pm 0.16 \times 10^{17}$ (the errors here obtained by adding in
quadrature
the errors for each), which agrees with
the total $N_{HI}$ of the whole absorber complex estimated from the
Lyman break.   This can
be seen qualitatively in Figure 2, from the good match of the fitted absorber
model at the
 Lyman Limit to
the regions of highest flux shortward of the limit.
The flux beyond the Lyman limit can be used to derive an   upper limit for the
total hydrogen column in the absorber complex, which yields
a fairly stringent lower limit on the $D/H$ ratio independent of HI line fits.
In order to try and circumvent any leftover instrument effects in this
part of the spectrum, the continuum of the QSO and the standard star were
assumed to have the same shape over this narrow range (50 \AA), and the shape
of the
standard star continuum was fit to match the parts of the QSO spectrum
at the red end of the order, in regions without conspicuous absorption.
This gave an
estimate for the continuum level, which was used both for the
high order line fit and the continuum level beyond the Lyman Limit,
and appears at the blue end of the fit in figure 2. The level of the signal
beyond the Lyman Limit was estimated in two ways. The simplest of these
used the mean of the counts per channel, blueward of the Lyman Limit in
this order. This gives an upper limit to the opacity of the cloud complex,
and hence an upper limit to the total hydrogen column of the cloud,
which can be used to find a very conservative lower limit to the $D/H$
ratio of the two absorbers. The second uses the mean of only those regions
blueward of the Lyman Limit where the counts per channel are at least 1
$\sigma$ above the zero-level, over a range of at least 1 \AA.
Two examples are the regions around 3932 \AA \
and 3941 \AA, each of which is a few \AA \ wide. Since additional
Ly-$\alpha$ and Ly-limit absorption must still occur
in these areas, the value for the opacity
determined from this estimate of the Lyman Limit signal is also an upper limit.
The two methods yield upper limits to the total HI column density of
$ 2.4 \times 10^{17}$ and $ 1.7 \times 10^{17} cm^{-2}$, respectively.

This limit almost exactly matches the sum of all the fitted components,
(including in particular number 16 in table 1 in addition to the two main
components),  so  there is almost no room to hide more than a small fraction
of additional HI anywhere in this complex beyond that already accounted for in
the fits.
Taking the lower limit on D to be $2.2 \times 10^{13} cm^{-2}$, which is
well constrained by the unsaturated $D\alpha$, therefore yields a firm lower
limit   $(D/H)\ge 1.3 \times 10^{-4}$, in good accordance with
the lower limits  determined from the hydrogen
line fits. Therefore, if the D candidate lines are indeed caused by
deuterium, $D/H$ must be high.

Most significantly for confirmation of deuterium, the DI lines are narrow,
making the interloper possibility very unlikely.
The fits for the two components are
$ 7.5 \pm 1.2 $ km/s and $ 8.8 \pm 1.1 $ km/s.
Hu et al (1995) find that in 3 QSOs (including this one, using the same data we
are
using, and others with similar resolution and signal-to-noise), only 11 out of
670
lines are found with $b\le 14$. This corresponds to less than 2\%, or only a
few lines
per QSO, in the 6000 resolution elements between
Ly$\alpha$ and Ly$\beta$, almost all of which are identified as metal lines.
Since interlopers must be metal lines, any coincidence must be purely
accidecntal,
(not affected by physical  Ly$\alpha$ autocorrelation), so we can compute the
interloper probability from Poisson statistics.
 The probability of a chance
coincidence of one of these lines at the precise velocity expected for
deuterium
is then much less than than $10^{-3}$ {\it for each component}.

Even more suggestive, a thermal model, likely to be appropriate
for such narrow lines,  predicts that the HI counterparts of
 the deuterium
lines should have $b=10.2\pm 1.7 {\rm km\ s^{-1}}$
and $ 12.4\pm 1.6 {\rm km\ s^{-1}}$, in agreement with the observed values of
$b=10.1\pm 2.0 {\rm km\ s^{-1}}$
and $ 12.7\pm 0.8 {\rm km\ s^{-1}}$. This coincidence also argues
against random interlopers.

In fact the HI lines are unusually narrow for forest lines, but this is not
surprising
since this system was selected for high column density, which often goes with
low ionization
parameter and temperature.
We also examined the data to look for metal absorption lines, but
 found only upper limits for CII, CIII, SiII and SiIII, all species
with several lines in regions of the spectrum which are unobstructed by
interlopers, and which have reasonable S/N.
These results are nevertheless consistent with a sensible physical model
of the main absorbing clouds. Using the models of Donahue and Shull (1991), we
adopt a conservative limit on the ionization parameter for the gas of $\log U
\buildrel <
\over \sim -4$ (where $U = n_{\gamma}/n_{H}$),
which gives an equilibrium temperature limit $T_4\le 1.2$
consistent with the narrow width of the H
and D features. This gives
 a neutral fraction of the hydrogen of ${\buildrel > \over \sim} 3
\times 10^{-2}$, and   estimated C and Si abundances
 ${\buildrel < \over \sim} 10^{-3}$ solar. With such a small ionization
parameter ($U << 10^{-2}$), charge-exchange reactions $D^{+} + H
\rightleftharpoons D + H^{+}$ guarantee that H and D are locked to the same
fractional ionization (J.\ Black, private communication), so
$N_{DI}/N_{HI}$ indeed gives the true deuterium abundance of the gas.

\section{Discussion}

 These results reinforce the interpretation of the deuterium
features in Q0014 as deuterium associated with the hydrogen.
First,
instead of one component we now have two cases where  the H and D
velocities agree.
A second coincidence is that both components
give the same   abundance -- that is, the estimated DI columns of the
individual components are  in the same ratio as
the columns of HI.
The third coincidence is that the  width of both DI features
 agrees with that  predicted for DI at the same temperature
as the hydrogen. Finally and most significantly, if the features
are interlopers they are too narrow to be part of the numerous
Ly$\alpha$ forest,
and therefore need to
be two of  the much rarer narrow and uncorrelated metal line population, making
the probability of two chance
interlopers at the precise required velocities
vanishingly small. By contrast,
the deuterium interpretation makes good physical sense, and
provides a natural explanation of all of these coinicidences.

The low metal abundance   makes it seem unlikely that more than a small
fraction of the gas has been processed through stars, so that a negligible
fraction of the initial deuterium has been destroyed.
Since the big bang is  the only known source of deuterium (\cite{rev73}),
we may take the absorber abundance  as our best estimate of
the  primordial abundance,
$ (D/H)_{p} = 1.9 \pm 0.4 \times 10^{-4}$.
Standard Big Bang Nucleosynthesis then gives a baryon to photon ratio
$\eta = 1.7 \pm 0.2 \times 10^{-10}$, corresponding to
$\Omega_bh^2 = 6.2 \pm 0.8 \times 10^{-3}$. For this value of $\eta$,
SBBN predictions are consistent with estimates of   cosmic
$^{4}$He and $^{7}$Li abundances (Copi et al. 1995ab, \cite{hat95},
\cite{dar}, \cite{fie96}).
The total baryon density is then only about three times larger than the
cosmic density of  known baryons in gas and stellar populations
(\cite{per92}), allowing a tidy picture where
 most cosmic baryons reside in or near galaxies
and clusters of galaxies rather than in a dominant,
diffuse intergalactic medium (\cite{fuk95}). There are not enough
extra baryons  to make  most of
the dark matter in galaxy halos, consistent with the constraints
from MACHO observations (\cite{alc95}), and implying that
the bulk of galactic dark matter is nonbaryonic.

Our estimated primordial deuterium abundance is   much larger than that of
the Milky
Way (\cite{lin95}), probably due to burning in stars,  which converts
it to
$^3$He. The low   value of $^3$He inferred  in the presolar nebula
and found in parts of
the interstellar medium (\cite{wil94}) can be explained as
further
conversion of the   $^{3}$He to heavier elements in stars (\cite{hog95},
\cite{was95}, \cite{cha95}, \cite{weiss96}, \cite{oli96}),
 rather than as a low primordial
value of deuterium.

\acknowledgments

We are particularly grateful to A.\ Songaila and L.\ L.\ Cowie for
performing the observations and sharing the data, R. Carswell for
use of his
VPFIT software and for help in using it, to the staff at NOAO, in
particular F.\ Valdez, for help with IRAF,    and to   J.\ Black,  S.\ Burles,
 C.\ Foltz, G.\ Wallerstein,
A.\ Wolfe  and the referee for helpful suggestions.
This work was supported at  the
University of Washington by NSF grant AST 932 0045.

\clearpage

\begin{deluxetable}{cccccccc}
\footnotesize
\tablecaption{ VPFIT results for the absorbers in Figure 1. \label{tbl-1}}
\tablewidth{0pt}
\tablehead{
\colhead{} &
\colhead{Ion} &
\colhead{logN} &
\colhead{$\pm$} &
\colhead{$z$} &
\colhead{$\pm$} &
\colhead{$b$ (km/s)} &
\colhead{$\pm$}
}
\startdata
 1 & HI & 15.45 & 0.68 & 2.310923 & 0.000015 & 22.1 & 4.3 \nl
 2 & HI & 13.18 & 0.09 & 2.644001 & 0.000052 & 25.0 & 6.5 \nl
 3 & HI & 13.22 & 0.11 & 2.648422 & 0.000038 & 16.8 & 5.7 \nl
 4 & HI & 13.71 & 0.04 & 2.649057 & 0.000025 & 24.7 & 3.0 \nl
 5 & HI & 12.78 & 0.34 & 3.315725 & 0.000309 & 34.3 & 12.4 \nl
 6 & HI & 13.24 & 0.12 & 3.316360 & 0.000029 & 17.6 & 3.5 \nl
 7 & HI & 14.01 & 0.07 & 3.317390 & 0.000050 & 25.3 & 2.9 \nl
 8 & HI & 14.28 & 0.05 & 3.318026 & 0.000028 & 19.6 & 2.2 \nl
 9 & HI & 12.99 & 0.42 & 3.318726 & 0.000388 & 38.0 & 12.9 \nl
 10 & HI & 13.74 & 0.34 & 3.320050 & 0.000064 &  10.4 & 5.0 \nl
 11 & DI & 13.03 & 0.10 & 3.320483 & 0.000010 &  7.5 & 1.2 \nl
 12 & HI & 16.76 & 0.07 & 3.320482 & 0.000044 & 10.1 & 2.0 \nl
 13 & DI & 13.18 & 0.07 & 3.320789 & 0.000012 &  8.8 & 1.1 \nl
 14 & HI & 16.90 & 0.06 & 3.320790 & 0.000037 & 12.7 & 0.8 \nl
 15 & HI & 15.22 & 0.38 & 3.321300 & 0.000131 & 13.1 & 4.3 \nl
 16 & HI & 16.36 & 0.16 & 3.322225 & 0.000041 & 22.8 & 2.1 \nl
 17 & HI & 15.36 & 0.19 & 3.322955 & 0.000088 & 17.8 & 6.0 \nl
 18 & HI & 14.44 & 0.32 & 3.323513 & 0.000089 & 12.0 & 5.1 \nl
 19 & HI & 13.06 & 0.13 & 3.324024 & 0.000078 & 24.0 & 6.4 \nl
\enddata
\end{deluxetable}

\clearpage

\centerline{\bf Figures}
{\parindent =0pc%\parskip =24pt

 Figure 1. \quad Spectral fits to Lyman $\alpha$, $\beta$, $\gamma$,
$\delta$, and $\zeta$, plotted on the same velocity scale.  Ticks mark the
centroids of individual  components, with parameters tabulated in Table 1;
the overall model fit comprising the sum of this absorption is the solid line.
The ticks corresponding to the bulk of the hydrogen column, and their deuterium
counterparts, are systems numbered 12/14 and 11/13, respectively.
[Postscript and GIF figures are available at
http://www.astro.washington.edu/rugers/res.html]

Figure 2. Spectral fit of the  same absorption model
to  Lyman 10 through 17. The data in the plot   (but not the  fit) has been
boxcar averaged at 7 km/sec to reduce noise. The plotted
model   does not include  photoelectric continuum
absorption and  recovers beyond the limit
to the QSO continuum; the predicted continuum from the line absorption model is
the
asymptotic value of the line fit curve longwards of the break, which agrees
with
the observed  flux beyond the limit.

}

\clearpage


\begin{thebibliography}{}

\bibitem[Alcock et al. 1995]{alc95}  Alcock, C., et al. 1995 {\it Phys.
Rev. Lett.}  {\bf 74}, 2867

\bibitem[Carswell et al. 1987]{car87} Carswell, R.F., Webb, J.K., Baldwin,
J.A., \& Atwood, B.\ (1987), {\it ApJ}, {\bf 319}, 709

\bibitem[Carswell et al. 1994]{car94}  Carswell, R F., Rauch, M., Weymann,
R. J., Cooke, A. J., and Webb, J. K., 1994, {\it MNRAS} {\bf 268}, L1

\bibitem[Chaffee et al. 1985]{cha85}  Chaffee, F. H., Foltz, C. B.,
R\"oser, H.-J., Weymann, R. J. \& Latham, D. W.\ 1985 {\it ApJ}, {\bf 292},
362

\bibitem[Chaffee et al. 1986]{cha86} Chaffee, F. H., Foltz, C. B., Bechtold, J.
\& Weymann, R. J.\ 1986, {\it ApJ}, {\bf 301}, 116

\bibitem[Charbonnel 1995]{cha95} Charbonnel, C. 1995, {\it ApJ}, {\bf 453},
L41.

\bibitem[Copi et al. 1995a]{cop95a} Copi, C. J., Schramm, D. N., and Turner,
M. S. 1995a, {\it Science} {\bf 267}, 192.

\bibitem[Copi et al. 1995b]{cop95b} Copi, C. J., Schramm, D. N., and Turner,
M. S. 1995b, {\it Phys.\
Rev.\ Lett.} {\bf 75}, 3981

\bibitem[Dar  1995]{dar} Dar, A. 1995, {\it ApJ}, {\bf 449}, 550

\bibitem[Donahue and Shull  1991]{doh91}  Donahue, M. \& Shull, J. M.\ 1991,
{\it ApJ}, {\bf 383}, 511.

\bibitem[Fields and Olive  1996]{fie96} Fields, B. D. and Olive, K. A. 1996,
Phys. Lett. B., in press.

\bibitem[Fukugita et al. 1995]{fuk95} Fukugita, M., Hogan, C. J. and
Peebles, P. J. E., 1996 {\it Nature}, in press.

\bibitem[Galli et al. 1995]{gal95}  Galli, D., Palla, F., Ferrini, F.\ and
Penco, U., 1995, {\it ApJ}, {\bf 443}, 536.

\bibitem[Hata et al. 1995]{hat95} Hata, N., Scherrer, R.\ J., Steigman, G.,
Thomas, D., Walker, T.\ P., Bludman, S., Langacker, P., 1995, {\it Phys.\
Rev.\ Lett.} {\bf 75}, 3977

\bibitem[Hogan 1995]{hog95} Hogan, C. J. \ 1995 {\it ApJ} {\bf 441}, L17.

\bibitem[Hu et al 1995]{hu} Hu, E. M., Kim, T.-S., Cowie, L. L.,
and Songaila, A. \ 1995, {\it AJ} {\bf 110}, 1526

\bibitem[Linsky et al. 1995]{lin95} Linsky, J.L., Diplas, A., Wood, B. E.,
Brown, A., Ayres, T. R.\ and Savage, B. D.  1995, {\it ApJ}, in press.

\bibitem[Olive et al. 1996]{oli96}  Olive, K. A., Rood, R. T., Schramm,
D. N., Truran, J. W., Vangioni-Flam, E.  1996, {\it ApJ}, submitted.

\bibitem[Persic and Salucci 1992]{per92}  Persic, M.\ and Salucci, P.
1992, {\it MNRAS}, {\bf 258}, 148.


\bibitem[Reeves et al. 1973]{rev73} Reeves, H., Audouze, J., Fowler, W.A.,
and Schramm, D.N. 1973, {\it ApJ}, {\bf 179}, 909.

\bibitem[Songaila, Cowie, Hogan and Rugers 1994]{schr} Songaila, A., Cowie,
L. L., Hogan, C. J., and Rugers, M. 1994, {\it Nature} {\bf 368}, 599.

\bibitem[Steigman 1994]{ste94} Steigman, G., 1994, {\it MNRAS} {\bf 269}, L53.

\bibitem[Tytler and Fan 1994]{tyt94} Tytler, D., and Fan, X.M. 1994,
Bull. of the AAS, Vol.\ 26 No.\ 4, 1424.

\bibitem[Wasserburg et al. 1995]{was95} Wasserburg, G. J., Boothroyd, A. I.,
Sackmann, I.-J. 1995, {\it ApJ Lett} {\bf 447}, L37.

\bibitem[Webb 1987]{web87} Webb, J.K. 1987, {\it PhD thesis}, Cambridge
University.


\bibitem[Weiss et al 1996]{weiss96} Weiss, A., Wagenhuber, J. and Denissenkov,
P. A. 1996,{\it A\&A}, in press.

\bibitem[Wilson and Rood 1994]{wil94}  Wilson, T. R., and Rood, R. T. 1994,
{\it ARA\&A}, {\bf 32}, 191.



\end{thebibliography}
\end{document}